\documentclass[preprint,onecolumn,nofootinbib]{revtex4}
\pdfoutput=1
\usepackage{CJK}
\usepackage[colorlinks=true,linkcolor=blue,urlcolor=blue,filecolor=black,citecolor=red,pdfstartview=FitV,pdftitle={},pdfsubject={},pdfkeywords={},pdfpagemode=None,bookmarksopen=true]{hyperref}
\usepackage{graphicx}
\usepackage{amsmath}
\usepackage{amsfonts}
\usepackage{amssymb,ulem}
\usepackage{color}%
\usepackage{dcolumn}
\newcommand{\f}{\begin{equation}}
\newcommand{\ff}{\end{equation}}
\newcommand{\fa}{\begin{eqnarray}}
\newcommand{\ffa}{\end{eqnarray}}

\begin{document}
\title{Holography of electrically and magnetically charged black branes}
\author{Zhenhua Zhou $^{1}$}
\thanks{dtplanck@163.com}
\author{Jian-Pin Wu $^{2,4}$}
\thanks{jianpinwu@yzu.edu.cn}
\author{Yi Ling $^{3,4,5}$}
\thanks{lingy@ihep.ac.cn}
\affiliation{$^1$~School of Physics and Electronic Information, Yunnan Normal University, Kunming, 650500, China}
\affiliation{$^2$~Center for Gravitation and Cosmology, College of Physical Science and Technology, Yangzhou University, Yangzhou 225009, China}
\affiliation{$^3$~Institute of High Energy Physics, Chinese Academy of Sciences, Beijing 100049, China}
\affiliation{$^4$~ Shanghai Key Laboratory of High Temperature Superconductors, Shanghai, 200444, China}
\affiliation{$^5$~School of Physics, University of Chinese Academy of Sciences, Beijing 100049, China}
\begin{abstract}
We construct a new class of black brane solutions in
Einstein-Maxwell-dilaton (EMD) theory, which is characterized by
two parameters $a,b$. Based on the obtained solutions, we make detailed
analysis on the ground state in zero temperature limit and find that for many cases it exhibits the behavior of
vanishing entropy density. We also find that the linear-T resistivity
can be realized in a large region of temperature for the specific
case of $a^2=1/3$, which is the Gubser-Rocha model dual to a
ground state with vanishing entropy density. Moreover, for $a=1$
we analytically construct the black brane which is magnetically
charged by virtue of the electric-magnetic (EM) duality. Various
transport coefficients are calculated and their temperature
dependence are obtained in the high temperature region.

\end{abstract}
\maketitle

\section{Introduction}
AdS/CFT correspondence provides a new direction for the study
of strongly correlated
systems\cite{Maldacena:1997re,Gubser:1998bc,Witten:1998qj,Aharony:1999ti}.
In particular, great progress has been made in modelling and
understanding the anomalous scaling behavior of the strange metal
phase (see \cite{Hartnoll:2015sea} and references therein).
Among of them the linear-T resistivity and quadratic-T
inverse Hall angle are two prominent properties of the
strangle metal, which have been widely observed in normal states
of high temperature superconductors as well as heavy fermion
compounds near a quantum critical point, which is universal
in a very wide range of temperature. By holography, the linear-T
resistivity has firstly been explored in
\cite{Hartnoll:2009ns,Davison:2013txa}. Then different scalings
between Hall angle and resistivity have also been investigated in
holographic framework
\cite{Pal:2010sx,Pal:2012gr,Gouteraux:2011xr,Kim:2010zq,Hoyos:2013eza,Gouteraux:2013oca,Gouteraux:2014hca,Lee:2011zzf,Lucas:2015pxa,Hartnoll:2015sea,Blake:2014yla}.
In particular, both the linear-T resistivity and quadratic-T
inverse Hall angle can be simultaneously reproduced in some
special holographic models
\cite{Karch:2014mba,Amoretti:2015gna,Blauvelt:2017koq,Zhou:2015dha,Chen:2017gsl}.

Currently it is still challenging to achieve the anomalous
scales of strange metal over a wide range of temperature in
holographic approach. It may be limited by the renormalization
group flow which is controlled by the specific bulk geometry
subject to Einstein field equations, and the scaling behavior of
the near horizon geometry of the background. Therefore, in this
direction one usually has two ways to improve the understanding of
the transport behavior of the dual system. One way is to consider
more general backgrounds within the framework of Einstein's
gravity theory. The other way is to introduce additional scaling
anomaly which may be characterized by Lifshitz dynamical exponent
and hyperscaling violating parameter. In the latter case, the
construction of asymptotic hyperscaling-violating and Lifshitz
solutions have largely improved the scaling analysis of the exotic
behavior in the strange metal
\cite{CiteHV1,CiteHV2,CiteHV3,CiteHV4,CiteHV5,CiteHV6,CiteHV7,CiteHV8,CiteHV9,CiteHV10}.
In this paper, we will focus on the former case, namely the
holographic construction of new backgrounds within the framework
of Einstein theory, without the involvement of scale anomaly. In
this way the Einstein-Maxwell-Dilaton (EMD) theory provides a nice
arena for the study of electric and magnetic transport phenomena in a strongly-coupled system. 
Previously a particular model constructed in EMD theory is the
Gubser-Rocha solution which describes an electrically charged
black brane\cite{Gubser:2009qt}. It is featured by a vanishing
entropy density at zero temperature\footnote{Another important
holographic model with vanishing entropy density ground
state is presented in \cite{Goldstein:2009cv}, in which the black
brane is numerically constructed and the near horizon geometry at
zero temperature possesses Lifshitz symmetry.}. This model
exhibits lots of peculiar properties similar to those of the
strangle metal, including the linear specific heat
\cite{Gubser:2009qt} and the linear resistivity at low temperature
\cite{Lucas:2014zea}. Also, as a typical model for holographic
studies, the Gubser-Rocha solution has been extended in various
circumstances, see e.g.
\cite{Donos:2014uba,CiteG-R2,CiteG-R3,CiteG-R4,CiteG-R5,CiteG-R6,Ling:2016wyr,Wu:2011cy}.

In this paper, we intend to construct new backgrounds which are
applicable for the study of both electric and magnetic transport
properties in holographic approach, aiming to provide more
comprehensive understanding on the anomalous behavior of strange
metals. We first analytically construct a new class of black brane
solutions which are electrically charged in
Einstein-Maxwell-dilaton (EMD) theory in Section \ref{GGb}. In
particular, we study the transport behavior in the dual system and
find that the linear-T resistivity holds in a large range of
temperature for $a^2=1/3$. Then by virtue of the electric-magnetic
(EM) duality for $a^2=1$, we construct a dyonic black brane
solution in Section \ref{dyonicbb} and Appendix
\ref{sec-EM-duality-bsol}. Various transport coefficients are
derived, including the resistivity, Hall angle, magnetic
resistance and Nernst coefficient. It is expected to provide a
useful platform for the study of both electric and magnetic
transport behavior in the holographic framework.

\section{Electrically charged dilatonic black branes}\label{GGb}

\subsection{Electrically charged dilatonic black branes}

We consider Einstein-Maxwell-Dilaton-Axion (EMDA) theory in four
dimensional spacetimes with the following action
\begin{equation}
S_{EMD}=\int d^4x \sqrt{-g}\Big(R-\frac{Z(\phi)}{4}F_{ab}F^{ab}-\frac{1}{2}(\partial \phi)^2+V(\phi)-\sum_{I=x,y}(\partial\psi_I)^2\Big)\,.\label{EMdac}
\end{equation}
The axionic fields $\psi_x,\,\psi_y$ are added to break the translation invariant,
which is responsible for the finite DC conductivity over a charged black hole background.

The black brane solutions of  EMDA theory and their holographic
properties have been widely studied in
\cite{Chan:1995fr,Cai:1996eg,Cai:1997ii,Charmousis:2009xr,Meyer:2011xn,Gouteraux:2011xr,Gouteraux:2011ce,Goldstein:2009cv,Gubser:2009qt,Ling:2013nxa,Caldarelli:2016nni,Gao:2005xv,Gao:2004tv,Gouteraux:2014hca}.
Analytical background can provide a more controllable pattern in
studying the holographic characteristics. To obtain an analytical
black brane solution, it is crucial to choose an appropriate
potential $V(\phi)$. For some specific form of $V(\phi)$,
the analytical AdS black brane solutions have been worked
out in
\cite{Gubser:2009qt,Gao:2005xv,Gao:2004tv,Gouteraux:2014hca,Caldarelli:2016nni}.
In this paper, we propose a more general form for the
potential and obtain a class of general AdS black brane solutions.
The potential $V(\phi)$ and the gauge coupling $Z(\phi)$ we choose
here are
\begin{subequations}
\label{Gpotential}
\begin{align}
&Z(\phi)=e^{a\phi}\,,\qquad V=V_0(1+\Phi)+V_1\,,\label{Zphi}\\
&V_0=\frac{6(a^2e^{\frac{\phi}{2a}}+e^{-\frac{a\phi}{2}})^2-2a^2(e^{\frac{\phi}{2a}}-e^{-\frac{a\phi}{2}})^2}{(1+a^2)^2}\,,\label{cp}\\
&V_1=\frac{2b}{(1+a^2)^2}(e^{\frac{1+a^2}{2a}\phi}-1)^3(a^2e^{-\frac{1}{a}\phi}+e^{-\frac{a^2+3}{2a}\phi})\,,\\
&\Phi=-b\Big(\frac{e^{\frac{3a^2-1}{2a}\phi}-1}{3a^2-1}+\frac{e^{\frac{a^2-3}{2a}\phi}-1}{a^2-3}-\frac{e^{\frac{a^2-1}{a}\phi}-1}{a^2-1}\Big)\,,\label{Gpot3}
\end{align}
\end{subequations}
where $a$ and $b$ are two free parameters in this EMDA
theory. And then, we take the following ansatz
\begin{subequations}
\label{ans}
\begin{align}
&ds^2=\frac{1}{z^2}\Big(-f(z)dt^2+\frac{dz^2}{f(z)}+(1+\Lambda z)^{\frac{2\beta^2}{1+\beta^2}}(dx^2+dy^2)\Big)\,,
\label{ans-f}
\\
&A=A_t(z)\,,~~~\phi=\phi(z)\,,~~~\psi_x=kx\,,~~~\psi_y=ky\,,
\label{ans-At}
\end{align}
\end{subequations}
Under the above setting, the EMDA theory \eqref{EMdac}, with the potential \eqref{Gpotential},
has two branches of the asymptotic AdS charged black brane solutions for $\beta=1/a$ and $\beta=a$, which are
\begin{subequations}
\label{Gsol1}
\begin{align}
\text{\textbf{Case 1:}~~~} &\beta=1/a\,,\\
&f(z)=(1+\Lambda z)^{\frac{2}{1+a^2}}h(z)\,,\\
&h(z)=1-\frac{q^2(1+a^2)}{4\Lambda}z^3(1+\Lambda z)^{-\frac{4}{1+a^2}}-k^2z^2(1+\Lambda z)^{\frac{a^2-3}{1+a^2}}\nonumber\\
&\qquad-b\Big(\frac{(1+\Lambda z)^{\frac{3a^2-1}{a^2+1}}-1}{3a^2-1}+\frac{(1+\Lambda z)^{\frac{a^2-3}{a^2+1}}-1}{a^2-3}-\frac{(1+\Lambda z)^{\frac{2a^2-2}{a^2+1}}-1}{a^2-1}\Big)\,,\\
&\phi(z)=\frac{2a}{1+a^2}\ln(\Lambda z+1)\,,~~~~A_t(z)=\mu-\frac{qz}{1+\Lambda z}\,,
\end{align}
\end{subequations}
\begin{subequations}
\label{Gsol2}
\begin{align}
\text{\textbf{Case 2:}~~~} &\beta=a\,,\\
&f(z)=(1+\Lambda z)^{\frac{2a^2}{1+a^2}}h(z)\,,\\
&h(z)=1+\frac{q^2(1+a^2)}{4\Lambda}z^3(1+\Lambda z)^{\frac{1-3a^2}{1+a^2}}-k^2z^2(1+\Lambda z)^{\frac{1-3a^2}{1+a^2}}\nonumber\\
&\qquad-b\Big(\frac{(1+\Lambda z)^{\frac{-3a^2+1}{a^2+1}}-1}{3a^2-1}+\frac{(1+\Lambda z)^{\frac{-a^2+3}{a^2+1}}-1}{a^2-3}-\frac{(1+\Lambda z)^{\frac{-2a^2+2}{a^2+1}}-1}{a^2-1}\Big)\,,\\
&\phi(z)=-\frac{2a}{1+a^2}\ln(\Lambda z+1)\,,\qquad A_t=\mu-qz\,.
\end{align}
\end{subequations}

Some remarks on these solutions are presented in what
follows.
\begin{itemize}
  \item The AdS boundary is located at $z=0$. $\mu,q$ are the chemical potential and charge density of the dual
boundary system, respectively.
 \item The parameter $\Lambda$ shall
be determined in terms of $a,b$ by the horizon condition
$h(z_+)=0$ with $z_+$ being the position of horizon. Namely,
only $a,b$ are free parameters in this model.
\item When
$b=0,a^2=1/3$, the solution of the case $\beta=1/a$ reduces to the
Gubser-Rocha one \cite{Gubser:2009qt}. When $b=0$, the solution of
the case $\beta=1/a$ becomes the well established results in
\cite{Gouteraux:2011xr,Gao:2005xv,Gao:2004tv,Gouteraux:2014hca}.
When~$k=q=0,b\neq0$, by redefining the parameters, the solution coincides to that in \cite{Astefanesei2014}.

\item To obtain the thermodynamics of the background, one need follow the standard holographic renormalization approach.
We would like to recommend article \cite{Astefanesei2015,Astefanesei2016,Caldarelli:2016nni},
in which the thermodynamics  of the above model with $b=0$ have been well studied\footnote{We are very grateful to Astefanesei for drawing our attention to \cite{Astefanesei2014}
as well as the correct holographic renormalization approach.}.
\end{itemize}

We would also like to point out that the above two branches of
solutions can be related by a coordinate transformation. That
is to say, the second branch of solutions can be obtained from the
first one under the following coordinate transformation
\begin{equation}
 z\rightarrow \frac{z}{1-\Lambda z}\label{tran}\,.
\end{equation}
Therefore, at this moment, it is enough to consider the first
branch with $\beta=1/a$ only.

Now, in order to make the solutions \eqref{Gsol1} become a black
brane background, one should consider the horizon condition
$h(z_+)=0$, which determines the location of the
horizon in terms of  $\Lambda$ and $a,b$ as below
\begin{align}
&1-\frac{q^2(1+a^2)}{4\Lambda}z_+^3(1+\Lambda z_+)^{-\frac{4}{1+a^2}}-k^2z_+^2(1+\Lambda z_+)^{\frac{a^2-3}{1+a^2}}\nonumber\\
&-b\Big(\frac{(1+\Lambda z_+)^{\frac{3a^2-1}{a^2+1}}-1}{3a^2-1}+\frac{(1+\Lambda z_+)^{\frac{a^2-3}{a^2+1}}-1}{a^2-3}-\frac{(1+\Lambda z_+)^{\frac{2a^2-2}{a^2+1}}-1}{a^2-1}\Big)=0\,.\label{Grel}
\end{align}
In addition, the derivative of $h(z)$ with respect to $z$ gives
\begin{align}
h'(z)=&-(1+\Lambda z)^{-\frac{4}{1+a^2}}\big(\frac{q^2z^2(3a^2-1)}{4\Lambda}+\frac{q^2z^2}{\Lambda(1+\Lambda z)}\nonumber\\
&+\frac{b\Lambda^3z^2}{1+a^2}+\frac{k^2z(3a^2-1)(1+\Lambda z)}{1+a^2}+\frac{k^2z(3-a^2)}{1+a^2}\big)\,.\label{Ghder}
\end{align}
From now on, we shall set $\Lambda>0,\,b\geq0$. When $k=0,q=0$,
$h(z)$ monotonically decreases from the boundary $h(z=0)=1$ to the
horizon, which guarantees $h(z)>0$ for the neutral black
brane background. When $q\neq0$, we can require
$a^2\geq1/3$ to ensure $h(z)>0$.

Then, the temperature and  entropy density can be calculated straightforward as
\begin{align}
T=&\frac{(1+\Lambda z_+)^{-\frac{2}{1+a^2}}}{4\pi }\Big(\frac{q^2z_+^2(3a^2-1)}{4\Lambda}+\frac{q^2z_+^2}{\Lambda(1+\Lambda z_+)}\nonumber\\
&+\frac{b\Lambda^3z_+^2}{1+a^2}+\frac{k^2z_+(3a^2-1)(1+\Lambda z_+)}{1+a^2}+\frac{k^2z_+(3-a^2)}{1+a^2}\Big)\,,\label{Gtep}\\
s=&\frac{4\pi}{ z^2_+}(1+\Lambda z_+)^{\frac{2}{1+a^2}}\,.\label{Gentropy}
\end{align}
It is obvious that the temperature $T>0$ when $h(z)$ is monotonic. This  is  the key point to
establish a zero temperature ground state with a zero entropy density, since if $T=0$ can be attained at a finite $z_+$, then, the  entropy density above must be finite.
On the contrary however, a positive $T>0$ for all finite $z_+$,  may decreases and becomes zero as $z_+\rightarrow 0$.

\subsection{Analysis on the ground state}\label{GGb-zero}

The ground state with zero entropy is physically acceptable. However, such ground state in holographic model is rare in the present literatures.
As we know, the only simple example is the  Gubser-Rocha solution \cite{Gubser:2009qt}. Now, with a more fruitful AdS background \eqref{Gsol1} at hand£¬
we give a detailed analysis, case by case, to find the ground state with a vanishing entropy density.
The method has been illustrated in the end of the last subsection, namely, we shall check whether $z_+\rightarrow\infty$ gives $T\rightarrow0$ as well as $s\rightarrow0$.

\subsubsection{Neutral black brane background for $q=0,k=0$}

We first consider the neutral black brane case without the axionic field, $q=0,\,k=0$. The horizon condition  \eqref{Grel}  and the temperature \eqref{Gtep} reduce to
\begin{align}
&b\Big(\frac{(1+\Lambda z_+)^{\frac{3a^2-1}{a^2+1}}-1}{3a^2-1}+\frac{(1+\Lambda z_+)^{\frac{a^2-3}{a^2+1}}-1}{a^2-3}-\frac{(1+\Lambda z_+)^{\frac{2a^2-2}{a^2+1}}-1}{a^2-1}\Big)=1\,,\\
&T=\frac{b\Lambda^3z_+^2}{4\pi(1+a^2)}(1+\Lambda z_+)^{-\frac{2}{1+a^2}}\,,
\end{align}
Then, we set $\Lambda=\Lambda_0/z_+$. When $z_+$ is varying, $\Lambda_0>0$ is a fixed parameter satisfying
\begin{align}
b\Big(\frac{(1+\Lambda_0)^{\frac{3a^2-1}{a^2+1}}-1}{3a^2-1}+\frac{(1+\Lambda_0)^{\frac{a^2-3}{a^2+1}}-1}{a^2-3}-\frac{(1+\Lambda_0)^{\frac{2a^2-2}{a^2+1}}-1}{a^2-1}\Big)=1\,,
\end{align}
Accordingly, the temperature and the entropy density \eqref{Gentropy} becomes
\begin{align}
T=\frac{b\Lambda_0^3}{4\pi z_+(1+a^2)}(1+\Lambda_0)^{-\frac{2}{1+a^2}}\,,\qquad s=\frac{4\pi}{ z^2_+}(1+\Lambda_0)^{\frac{2}{1+a^2}}\,.
\end{align}
When $z_+$ is varying, we have a simple relation
\begin{align}
s\propto T^2\,.
\end{align}
Both the $s,T$ tends to zero as $z_+\rightarrow\infty$. Such neutral background admits a ground state with zero entropy density.

\subsubsection{Simple charged black brane background for $b=0$}

Next, we study a simple charged black brane case with $q\neq0$ but $b=0$, where the general solution \eqref{Gsol1} reduces to
\begin{subequations}
\label{sol1}
\begin{align}
&f=(1+\Lambda z)^{\frac{2}{1+a^2}}h(z)\,,~~~h=1-\frac{q^2(1+a^2)}{4\Lambda}z^3(1+\Lambda z)^{-\frac{4}{1+a^2}}-k^2z^2(1+\Lambda z)^{\frac{a^2-3}{1+a^2}}\,,
\label{sol1f}
\
\\
&A_t=\mu-\frac{qz}{1+\Lambda z}\,,~~~~\phi=\frac{2a}{1+a^2}\ln(\Lambda z+1)\,,
\label{sol1At}
\end{align}
\end{subequations}
The above solutions have been well studied in \cite{Gao:2005xv,Gao:2004tv,Gouteraux:2014hca,Caldarelli:2016nni}. Especially, when $a^2=1/3$, it reduces the famous Gubser-Rocha solution \cite{Gubser:2009qt}.
Here, we try to complete the zero temperature analysis. We will find that not only the Gubser-Rocha solution, i.e., $a^2=1/3$, but also the case of $a^2>3,\,k=0$, are the vanishing entropy density background at the zero temperature.

We first consider the case without axion fields, namely $k=0$ where the horizon condition  \eqref{Grel} and the
temperature \eqref{Gtep} reduces to
\begin{align}
&\frac{q^2(1+a^2)}{4\Lambda}z_+^3(1+\Lambda z_+)^{-\frac{4}{1+a^2}}=1\,,\label{b0rel}\\
&T=\frac{(1+\Lambda z_+)^{-\frac{2}{1+a^2}}}{4\pi }\Big(\frac{q^2z_+^2(3a^2-1)}{4\Lambda}+\frac{q^2z_+^2}{\Lambda(1+\Lambda z_+)}\Big)\,.\label{b0k0tep}
\end{align}
When $a^2<1/3$, $T=0$ can be achieved at a finite position
\begin{eqnarray}
z_+=\frac{1}{\mu}\sqrt{\frac{3}{4}\Big(\frac{4}{1-3a^2}\Big)^{\frac{3-a^2}{1+a^2}}}\,.
\end{eqnarray}
The corresponding entropy density is
\begin{eqnarray}
s=\frac{\mu^2}{3\pi}\Big(\frac{4}{1-3a^2}\Big)^{\frac{a^2-1}{1+a^2}}\,,
\end{eqnarray}
which is of course finite.

However, when $a^2\geq1/3$, the story is totally different,
because the temperature \eqref{b0k0tep} is always positive. We then check the  temperature behavior as
$z_+\rightarrow\infty$.

In the limit of $z_+\rightarrow\infty$,
the relation \eqref{b0rel} forces that $\Lambda z_++1\approx\Lambda z_+ \rightarrow\infty$ and gives
\begin{eqnarray}
\Lambda=\Big(\frac{(1+a^2)q^2}{4}\Big)^{\frac{a^2+1}{a^2+5}}z_+^{\frac{3a^2-1}{a^2+5}}\sim z_+^{\frac{3a^2-1}{a^2+5}}\rightarrow\infty\,.
\end{eqnarray}
Then the temperature in \eqref{b0k0tep} and the entropy density
in  \eqref{Gentropy} approximatively read as
\begin{eqnarray}
T=\frac{3a^2-1}{4\pi(1+a^2)}\Big(\frac{(1+a^2)q^2}{4}\Big)^{\frac{2}{a^2+5}}z_+^{\frac{3-a^2}{a^2+5}}\,,\qquad s=4\pi\Big(\frac{(1+a^2)q^2}{4}\Big)^{\frac{2}{a^2+5}}z_+^{-\frac{2a^2+2}{a^2+5}}\,.
\end{eqnarray}
It is clear that for $a^2\geq1/3$ the entropy density deceases to
zero as $z_+\rightarrow\infty$. But the zero temperature and zero
entropy density can be simultaneously achieved only for
$a^2=1/3$ or $a^2>3$. On the contrary, no zero
temperature exists for $3\geq a^2>1/3$. Actually,  as
$z_+\rightarrow\infty$, it is a high temperature limit with
$T\rightarrow\infty$.

Next, we consider the case $k\neq0$.  The  horizon condition  \eqref{Grel}  and the temperature \eqref{Gtep} with $b=0$ reduces to
\begin{align}
&1-\frac{q^2(1+a^2)}{4\Lambda}z_+^3(1+\Lambda z_+)^{-\frac{4}{1+a^2}}-k^2z_+^2(1+\Lambda z_+)^{\frac{a^2-3}{1+a^2}}=0\,,\label{Lambda-a}\\
&T=\frac{(1+\Lambda z_+)^{-\frac{2}{1+a^2}}}{4\pi }\Big(\frac{q^2z_+^2(3a^2-1)}{4\Lambda}+\frac{q^2z_+^2}{\Lambda(1+\Lambda z_+)}\nonumber\\
&\qquad+\frac{k^2z_+(3a^2-1)(1+\Lambda z_+)}{1+a^2}+\frac{k^2z_+(3-a^2)}{1+a^2}\Big)\,,\label{b0tep}
\end{align}
Then, taking the  limit $z_+\rightarrow\infty$, the horizon condition \eqref{Lambda-a} becomes
\begin{eqnarray}
1-\frac{q^2(1+a^2)}{4\Lambda }z_+^3(\Lambda z_+)^{\frac{-4}{1+a^2}}-k^2z_+^2(\Lambda z_+)^{\frac{a^2-3}{1+a^2}}=0\,.\label{Lambda-aapp}
\end{eqnarray}
Immediately, we find that
one can not obtain an extremal black brane solution with axionic
fields for $a^2\geq 3$. It is in contrast to the case without
axionic fields.
Therefore, we conclude that once the axionic fields are taken
into account, the simple charge black brane solution with vanishing ground
state entropy density can be achieved only for $a^2=1/3$.

\subsubsection{Special cases for $a^2=1,a^2=1/3,a^2=3$}

So far, the discussion is based on the general potential
\eqref{Gpotential}. For the special cases
$a^2=1,a^2=1/3,a^2=3$, it is convenient to write down the form of
$\Phi$ in the potential \eqref{Gpotential} after taking the limit,
which are separately given by
\begin{subequations}
\begin{align}
&\Phi(a^2=1/3)=-b\Big(\frac{\sqrt{3}\phi}{2}-\frac{3e^{-\frac{4\sqrt{3}}{3}\phi}-3}{8}+\frac{3e^{-\frac{2\sqrt{3}}{3}\phi}-3}{2}\Big)\,,\\
&\Phi(a^2=1)=-b\Big(\sinh \phi-\phi\Big)\,,\\
&\Phi(a^2=3)=\frac{-b}{24} \left(3 e^{\frac{2 \phi }{\sqrt{3}}} \left(e^{\frac{2 \phi }{\sqrt{3}}}-4\right)+4 \sqrt{3} \phi +9\right)\,,
\end{align}
\end{subequations}
The background solutions of $h(z)$ in \eqref{Gsol1} also take
the form as
\begin{subequations}
\label{Sprel}
\begin{align}
&h(a^2=\frac{1}{3})=1-\frac{q^2z^3}{3\Lambda(1+\Lambda z)^3}-\frac{k^2z^2}{(1+\Lambda z)^2}-b\left(\frac{3}{4} \log (1+\Lambda z)-\frac{3 \Lambda  z (3 \Lambda  z+2)}{8 (\Lambda  z+1)^2}\right)\,,\\
&h(a^2=1)=1-\frac{q^2z^3}{2\Lambda(1+\Lambda z)^2}-\frac{k^2z^2}{(1+\Lambda z)}-b\left(\frac{\Lambda  z (\Lambda  z+2)}{2 \Lambda  z+2}-\log (1+\Lambda z)\right)\,,\\
&h(a^2=3)=1-\frac{q^2z^3}{\Lambda(1+\Lambda z)}-k^2z^2-b\left(\frac{1}{8} (\Lambda z (\Lambda z-2)+2 \log (1+\Lambda z))\right)\,,
\end{align}
\end{subequations}
However, the special parameters would not change the formula of
$h'(z)$  and hence the temperature formula. Namely, we can use the
former expressions \eqref{Ghder} and  \eqref{Gtep}  directly and
fix $a^2=1,a^2=1/3,a^2=3$ respectively.

When we do the zero temperature analysis, the only thing needing to change is the the horizon condition \eqref{Grel}, which should be replaced by using the above results \eqref{Sprel} and require
$h(a^2=1/3,z_+)=0,\,h(a^2=1,z_+)=0,\,h(a^2=3,z_+)=0$ for $a^2=1,a^2=1/3,a^2=3$ respectively.

It is easy to see that, for the neutral case $q=0,k=0$,  a zero temperature background with vanishing entropy also exists for these special parameters.
While, for $q=0,k\neq0$, due to the logarithm divergence as $z_+\rightarrow\infty$ in \eqref{Sprel}, the system  do not have ground state with vanishing entropy.
For the simple charge case $b=0,k=0$, special parameters have no influence on the previous discussion. We still have the zero entropy background at zero temperature for $a^2=1/3$ or $a^2>3$.

\section{Linear-T resistivity}

In this section, we consider the electric transport behavior of the
dual system over the black brane geometry \eqref{Gsol1}.
Specifically, we calculate the DC conductivity with the
interest in its dependence on the temperature. We find the Gubser-Rocha case exhibits a linear-T resistivity valid in a wide temperature, which coincides to the universal behaviors of the strange metal.

Before the discussion, we should scale the quantities by the charge density $q$. Namely, in this section
$z_+,\,\Lambda,\,k,\,T$ should be understood as the scaled ones $\sqrt{q}z_+,\,\Lambda/\sqrt{q},\,k/\sqrt{q},\,T/\sqrt{q}$.
Then, using the standard
holographic techniques, one can derive the DC conductivity as
\begin{equation}
\sigma=(1+\Lambda z_{+})^{\frac{2a^2}{1+a^2}}\Big(1+\frac{z_+^2}{2k^2(1+\Lambda z_+)^2}\Big)\,.\label{dccond}
\end{equation}
Usually the relation between $\sigma$ and $T$ is
complicated, since one needs to first  solve $z_+(T),\,\Lambda(T)$ by the relations in Eqs. \eqref{Grel} and \eqref{Gtep}, and then substitute them into the above expression of the conductivity.
It is thus hard to catch a simple universal relation of the resistivity of the temperature.
To simplify, one need take some approximation. We firstly consider the `deep horizon limit' $z_+\rightarrow\infty$ as appeared in lots of the literature. Recall that when $b\neq0$, such limit is invalid for the divergence in the horizon conditions, see for instance \eqref{Sprel}.
Thus, we only consider the simple charge black brane case with $b=0$.

In this
limit $z_+\rightarrow\infty$, which corresponds to
low temperature limit at $a^2=1/3$ or high temperature
limit for $3>a^2>1/3$, the expression of temperature in \eqref{b0tep} can be
largely simplified. Together with the relation
\eqref{Lambda-aapp}, we have
\begin{subequations}p
\label{sigma}
\begin{align}
&\sigma\sim T^2\,,\qquad\qquad3>a^2>1/3\,,\\
&\sigma\sim 1/T\,,\qquad\qquad a^2=1/3\,.
\end{align}
\end{subequations}
The result indicates that for $1/3<a^2<3$, the resistivity decrease as $1/T^2$
in high temperature limit, which is independent of $a$. While for
$a^2=1/3$, this holographic system captures the important property
of the strange metal, i.e., linear-T resistivity \cite{Davison:2013txa}.
However, the behavior of the linear-T resistivity only holds in low temperature limit in \cite{Davison:2013txa}.
Here we shall address that the linear-T resistivity can also holds for a large temperature range if $\Lambda\gg 1$ as well as $\Lambda z_+\gg1$.

We firstly check whether such region exists. When $a^2=1/3$, the horizon
condition \eqref{Lambda-a} becomes
\begin{align}
1-\frac{1}{3\Lambda}z_+^3(1+\Lambda z_+)^{-3}-k^2z_+^2(1+\Lambda z_+)^{-2}=0\,.\label{1/3hc}
\end{align}
To be clear, we rewrite the condition as
\begin{align}
1-\frac{1}{3\Lambda^4}y^3-\frac{k^2}{\Lambda^2}y^2=0\;\Leftrightarrow\; 2\Lambda^2=k^2y^2+\sqrt{k^4y^4+4y^3/3}\,,
\end{align}
where $y=(1+1/(\Lambda z_+))^{-1}$. We can see that $\Lambda z_+\gg1$ leads to
$y>1/2$ and then
\begin{align}
2\Lambda^2>k^2/2+\sqrt{k^4/8+1/6}>k^2/2\,.
\end{align}
Thus, when $k\gg1$  the regime with $\Lambda\gg 1,\Lambda
z_+\gg1$ survives.

\begin{figure}
\center{
\includegraphics[scale=0.5]{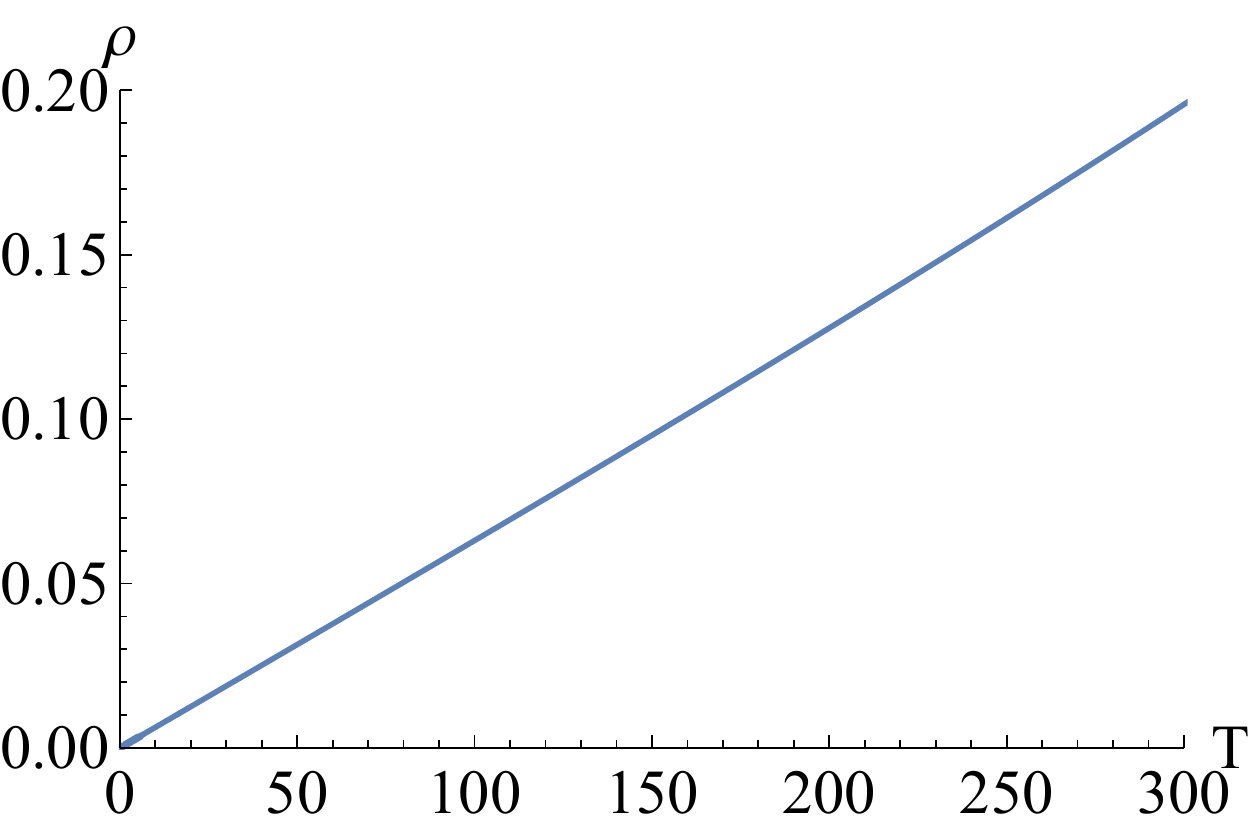}\ \hspace{0.8cm}\ \\
\caption{\label{rvstem} The linear-T resistivity behavior in a large temperature range. Here, we have set $k=10^4$, which ensures $k\gg1$, $\Lambda\gg 1$ and $\Lambda z_+\gg1$.}}
\end{figure}
Now we turn to explore the relation between the resistivity
and the temperature. Using Eqs. \eqref{TandDC-linear} and
\eqref{1/3hc}, we plot the resistivity as the function of the
temperature in FIG.\ref{rvstem}. Obviously, the resistivity
linearly depends on the temperature. Especially, we can see that
the linear-T resistivity survives in a large range of
temperature, which shall be further addressed in what
follows.

Also, we can obtain an approximate expression of the
resistivity. When $a^2=1/3$, both the temperature in
\eqref{b0tep} and the conductivity in \eqref{dccond} reduce
to
\begin{equation}
T=\frac{1}{4\pi\sqrt{z_+\Lambda}}(2\Lambda+\frac{1}{3\Lambda^3})\,,\qquad \sigma=\sqrt{z_+\Lambda}(1+\frac{1}{2\Lambda^2k^2})\,.
\label{TandDC-linear}
\end{equation}
When $k\gg1$ and $\Lambda\gg 1,\Lambda z_+\gg1$, the above equation can be approximately expressed as
\begin{align*}
T=\frac{1}{4\pi\sqrt{z_+\Lambda}}(2\Lambda+\frac{1}{3\Lambda^3})\approx \frac{1}{4\pi\sqrt{z_+k}}(2k+\frac{1}{3k^3})\,.
\end{align*}
Note that in the above equation, we have used the relations,
$y\rightarrow1$ and $\Lambda\rightarrow k$, which can be directly
derived from the condition of $\Lambda z_+\gg1$. Obviously, $T$ is
not sensitive to $\Lambda$, but mainly depends on $z_+$. The
situation for the conductivity is similar. Both $T$ and
$\sigma$ change mainly with $z_+$, while $\Lambda$ can be treated
as a constant, i.e. $\Lambda=k$. And then, with the use of
the relation $k\gg1$, the temperature and the conductivity can be
further simplified as
\begin{equation}
T\approx\frac{1}{2\pi}\sqrt{\frac{k}{z_+}}\,,\qquad \sigma\approx\sqrt{z_+k}\,.\label{TandDC-linear1}
\end{equation}
Immediately, we obtain a linear-T resistivity law
\begin{equation}
\rho\approx\frac{2\pi}{k} T\,.
\end{equation}

\begin{table}[ht]
\begin{tabular}{c|c|c|c}
  $k=10^4$~&~$z_+=10^{-3}$~&~$z_+=10^{-2}$~&~$z_+=10^2$\\
   \hline   \hline
  $\Lambda$~&~$9000.00$~&~$9900.00$~&~$9999.99$\\
  $T$~&~$477.47$~&~$158.36$~&~$1.59$\\
    \hline
      \hline
  $\Lambda$~&~$\approx10^4$~&~$\approx10^4$~&~$\approx10^4$\\
   $T$~&~$\approx503.29$~&~$\approx159.16$~&~$\approx1.59$\\
\end{tabular}\qquad\qquad
\begin{tabular}{c|c|c}
  $k=10^5$~&~$z_+=10^{-2}$~&~$z_+=10^2$\\
   \hline   \hline
  $\Lambda$~&~$99899.99$~&~$99999.99$\\
  $T$~&~$503.04$~&~$5.03$\\
    \hline
      \hline
  $\Lambda$~&~$\approx10^5$~&~$\approx10^5$\\
   $T$~&~$\approx503.29$~&~$\approx5.03$\\
\end{tabular}
\caption{\label{tableI} The exact value of $\Lambda$ and the temperature $T$ obtained by the expressions in \eqref{1/3hc} and \eqref{TandDC-linear}, which is presented in the second and third rows,
and the approximate results obtained in terms of the first equation in Eq.\eqref{TandDC-linear1}, which is presented in the fourth and fifth rows.}
\end{table}

Next, we shall address that the linear-T resistivity survives
in a large range of temperature. To this end, we present
two examples as what follows (see TABLE \ref{tableI}). The exact value of $\Lambda$ and the temperature $T$ in TABLE \ref{tableI} are calculated by the
expressions in \eqref{1/3hc} and \eqref{TandDC-linear}. The approximate results of the
temperature are obtained in terms of the first equation in
Eq.\eqref{TandDC-linear1}. Note that we have used the relation
$\Lambda\thickapprox k$. From the above table (the second and third rows),
we confirm the result that when $k\gg1$, $\Lambda\gg 1$, and
$\Lambda z_+\gg1$, $\Lambda$ is approximately a constant, i.e.,
$\Lambda\thickapprox k$. In addition, the approximate values of the temperature are consistent with the exact ones,
which means that the approximate linear-T
resistivity expression, namely Eq. \eqref{TandDC-linear1},
holds very well in the region of $k\gg1$, $\Lambda\gg 1$, and
$\Lambda z_+\gg1$. More importantly, from this table, we can see
that the temperature $T$ indeed varies in terms of $z_+$, crossing
a large range.

In summary, the linear-T resistivity is achieved when
$\Lambda\approx k\gg1$, which is a good approximation. It always
happens if one considers a system with large parameter
$k\gg1$. In our  examples, $T<160$ is the  good regime of linear-T
resistivity  for $k=10^4$, while $T<503$ is, at least, the good
regime for $k=10^5$. Thus, for the parameters satisfying the
certain conditions, the linear-T resistivity is achieved for a
wide range of temperature.

\textbf{Note added.} \emph{As this work was being completed, we
were informed from Chao Niu  that they also find the
linear-T resistivity behavior at high temperature region in
\cite{Jeong:2018tua}.}

\section{Dyonic dilatonic black brane and its Transports}\label{dyonicbb}
The lack of exact dyonic solution in gravity theory prevents us
from investigating the magnetic transport behavior of the dual
system in an analytical manner. Fortunately, in EMDA theory
\eqref{EMdac}, we are able to find such an analytical dyonic
solution for the special case of $a^2=1$ by virtue of the
electromagnetic self-duality. Here we just list the dyonic
solutions as below. The detailed derivation can be found in
Appendix \ref{sec-EM-duality-bsol}. Moreover, we point out that an
AdS dyonic solution with $b=0$ as well as $k=0$ has previously
been reported in \cite{dyonic2013}\footnote{We are very
grateful to Gouter\'{a}ux for drawing our attention to the work in
\cite{Caldarelli:2016nni,dyonic2013}.}. In
\cite{Caldarelli:2016nni}, by detailed analysis on the boundary
condition it is argued that a dyonic solution may only exist at
$a=1$ ( $\xi=1$ in their paper). Here, interestingly enough, we
provide an interpretation for this fact from a different angle of
view, namely the electromagnetic self-duality.

%

For $a=1$, the EMDA theory with the potential \eqref{Gpotential} exists the dyonic black brane solution as the following
form
\begin{subequations}
\label{dyonic-axions-sol}
\begin{align}
&ds^2=\frac{1}{z^2}(-f(z)dt^2+\frac{dz^2}{f(z)}+(1+\Lambda z)(dx^2+dy^2))\,,
\label{das-ds}\\
&\phi=\ln(\Lambda z+1)\,,\label{das-phi}\\
&A=\Big(\mu-\frac{qz}{1+\Lambda z}\Big)dt+Bxdy\,,\label{das-A}\\
&f=(1+\Lambda z)h(z)\,,\label{das-f}\\
&h=1-\frac{q^2}{2\Lambda}z^3(1+\Lambda
z)^{-2}+\frac{B^2}{2\Lambda}z^3(1+\Lambda z)^{-1}-k^2z^2(1+\Lambda
z)^{-1}\nonumber\\
&\qquad-b\big(\frac{\Lambda z (\Lambda z+2)}{2(1+\Lambda z)}-\log (1+\Lambda z)\big)\label{das-h}\,,
\end{align}
\end{subequations}
where $B$ is a constant magnetic field. The horizon
condition $h(z_+)=0$ gives rise to
\begin{align}
&1-\frac{q^2}{2\Lambda}z_+^3(1+\Lambda
z_+)^{-2}+\frac{B^2}{2\Lambda}z_+^3(1+\Lambda z_+)^{-1}-k^2z_+^2(1+\Lambda
z_+)^{-1}\nonumber\\
&\qquad-b\big(\frac{\Lambda z_+ (\Lambda z_++2)}{2(1+\Lambda z_+)}-\log (1+\Lambda z_+)\big)=0\label{Gdyrel}\,,
\end{align}
The temperature of this black
brane is
\begin{align}
T=&\frac{(1+\Lambda z_+)^{-1}}{4\pi }\Big(\frac{q^2z_+^2}{2\Lambda}+\frac{q^2z_+^2}{\Lambda(1+\Lambda z_+)}+k^2z_+(1+\Lambda z_+)+k^2z_+\nonumber\\
&+\frac{b\Lambda^3z_+^2}{2}-\frac{B^2z_+^2}{2\Lambda}-\frac{B^2z_+^2}{\Lambda}(1+\Lambda z_+)\Big)\,,\label{Gdytep}
\end{align}

Now, we turn to study the DC transports. Employing the
standard techniques developed in
\cite{Donos:2014uba,Donos:2014cya,Blake:2014yla}, we obtain the
thermoelectric conductivities over the dyonic black brane
geometry \eqref{dyonic-axions-sol} as
\begin{subequations}
\begin{align}
&\sigma_{xx}=\frac{H(q^2+HZ+B^2Z^2)}{B^2q^2+(B^2Z+H)^2}\,,~~~~~\sigma_{xy}=\frac{Bq(q^2+2HZ+B^2Z^2)}{B^2q^2+(B^2Z+H)^2}\,,\\
&\alpha_{xx}=\frac{Hsq}{B^2q^2+(B^2Z+H)^2}\,,~~~~~\alpha_{xy}=\frac{Bs(q^2+HZ+B^2Z^2)}{B^2q^2+(B^2Z+H)^2}\,,\\
&\overline{\kappa}_{xx}=\frac{s^2T(B^2Z+H)}{B^2q^2+(B^2Z+H)^2}\,,~~~~~\overline{\kappa}_{xy}=\frac{Bqs^2T}{B^2q^2+(B^2Z+H)^2}\,,
\end{align}
\end{subequations}
where $Z\equiv Z(\phi)|_{z_+}$, $H\equiv H(\phi)|_{z_+}$ and $q$ are, respectively
\begin{align}
Z=(1+\Lambda z_+)\,,~~~~H=\frac{2k^2(1+\Lambda z_+)}{z^2_+}\,,~~~~q=\frac{(1+\Lambda z_+)\mu}{z_+}\,.
\end{align}
And then, we give the charge transport coefficients, the DC resistivity $\rho_{dc}$ and the thermopower $S$ as
\begin{subequations}
\begin{align}
&\rho_{dc}=\frac{1}{\sigma_{xx}(B=0)}=\frac{2 k^2}{(1+\Lambda z_+)(2 k^2+\mu ^2) }\,,\\
&S=\frac{\alpha_{xx}(B=0)}{\sigma_{xx}(B=0)}=\frac{4 \pi  \mu }{z_+(2k^2 +\mu ^2) }\,.
\end{align}
\end{subequations}
Also the magnetic transport coefficients, including the Hall angle $\tan\theta_H$, the Hall Lorentz ratio $L_H$, the magneticresistance $\rho_{B}$ and Nernst coefficient $\nu$ are calculated explicitly as
\begin{subequations}
\begin{align}
&\tan\theta_H=\frac{\sigma_{xy}}{\sigma_{xx}}=\frac{B \mu  z \left(B^2 z^2+4 k^2+\mu ^2\right)}{2 k^2 \left(B^2 z^2+2 k^2+\mu ^2\right)}\,,\\
&L_H=\frac{\overline{\kappa}_{xy}}{T\sigma_{xy}}=\frac{16 \pi ^2}{z^2 \left(B^2 z^2+4 k^2+\mu ^2\right)}\,,\\
&\rho_{B}=\frac{\rho_{xx}-\rho_{xx}(B=0)}{\rho_{xx}(B=0)}=\frac{2 B^2 k^2 z^2}{\mu ^2 \left(B^2 z^2+4 k^2\right)+4 k^4+\mu ^4}\,,\\
&\nu=\frac{1}{B}(\frac{\alpha_{xy}}{\sigma_{xx}}-S\tan\theta_H)=\frac{4 \pi  \left(B^2 z^2+2 k^2\right)}{\left(2 k^2+\mu ^2\right) \left(B^2 z^2+2 k^2+\mu ^2\right)}\,.
\end{align}
\end{subequations}
In order to simplify the expression of temperature, we
will only consider the case $b=0$ in the follows, in which
we can take a large $z_+$ limit. The horizon condition
\eqref{Gdyrel} and the temperature  \eqref{Gdytep} reduce to
\begin{align}
&1-\frac{q^2}{2\Lambda}z_+^3(1+\Lambda z_+)^{-2}+\frac{B^2}{2\Lambda}z_+^3(1+\Lambda z_+)^{-1}-k^2z_+^2(1+\Lambda z_+)^{-1}=0\,,\label{dyrelb0}\\
&4\pi T=(1+\Lambda z_+)\Big(\frac{3}{z_+}-\frac{\Lambda}{1+\Lambda z_+}-\frac{q^2z^3_+}{2(1+\Lambda z_+)^3}-\frac{k^2z_+}{1+\Lambda z_+}\Big)\,.\label{dytepb0}
\end{align}
We are particularly interested in the transport
behavior of this system in the high temperature limit,
in which $\Lambda z_+\gg 1$ such that $1+\Lambda z_+\approx
\Lambda z_+$. In this case, Eqs. (\ref{dyrelb0}) and \eqref{dytepb0}
reduce to
\begin{subequations}
\begin{align}
&1-\frac{q^2}{2\Lambda^3}z_+-\frac{k^2}{\Lambda}z_++\frac{B^2}{2\Lambda^2}z_+^2=0\,,\\
&4\pi T=\Lambda\Big(2-\frac{q^2}{2\Lambda^3}z_+-\frac{k^2}{\Lambda}z_+\Big)\,.
\end{align}
\end{subequations}
Next we consider the situation that the magnetic field $B$
is small, then $z_+$ and $\Lambda$ can be solved as
\begin{subequations}
\begin{align}
&\Lambda=4 \pi  T+\frac{2048 \pi ^5  T^5}{\left(32 \pi ^2 k^2 T^2+q^2\right)^2}B^2+\mathcal{O}(B^4)\,,\\
&z_+=\frac{128 \pi ^3 T^3}{32 \pi ^2 k^2 T^2+q^2}+\frac{262144 \pi ^7  T^7 \left(16 \pi ^2 k^2 T^2+q^2\right)}{\left(32 \pi ^2 k^2 T^2+q^2\right)^4}B^2+\mathcal{O}(B^4)\,.
\end{align}
\end{subequations}
In the above equations, we have expressed $z_+$ and $\Lambda$ up
to the second order of $B$. As a consequence, we give the
charge and magnetic transport coefficients up to the first order
as
\begin{subequations}
\begin{align}
&\rho_{dc}=\frac{k^2}{16 \pi ^2 T^2}\,,\qquad S=\frac{q}{8 \pi  T^2}\,,\\
&\tan\theta_H=B\frac{16 \pi ^2  q T^2 \left(64 \pi ^2 k^2 T^2+q^2\right)}{k^2 \left(32 \pi ^2 k^2 T^2+q^2\right)^2}+\mathcal{O}(B^3)\,,\\
&L_H=\frac{16 \pi ^2 k^4}{64 \pi ^2 k^2 T^2+q^2}+\frac{q^2}{64 \pi ^2 T^4}+\mathcal{O}(B^2)\,,\\
&\rho_{B}=- B^2 \frac{512\left(\pi ^4 T^4 \left(192 \pi ^2 k^2 q^2 T^2+1024 \left(3 \pi ^4 k^4 T^4-16 \pi ^6 k^2 T^6\right)+3 q^4\right)\right)}{\left(32 \pi ^2 k^2 T^2+q^2\right)^4}+\mathcal{O}(B^4)\,,\\
&\nu=\frac{2048 \pi ^5 k^2 T^4}{\left(32 \pi ^2 k^2 T^2+q^2\right)^2}+\mathcal{O}(B^2)\,.
\end{align}
\end{subequations}

\begin{figure}
\center{
\includegraphics[scale=0.5]{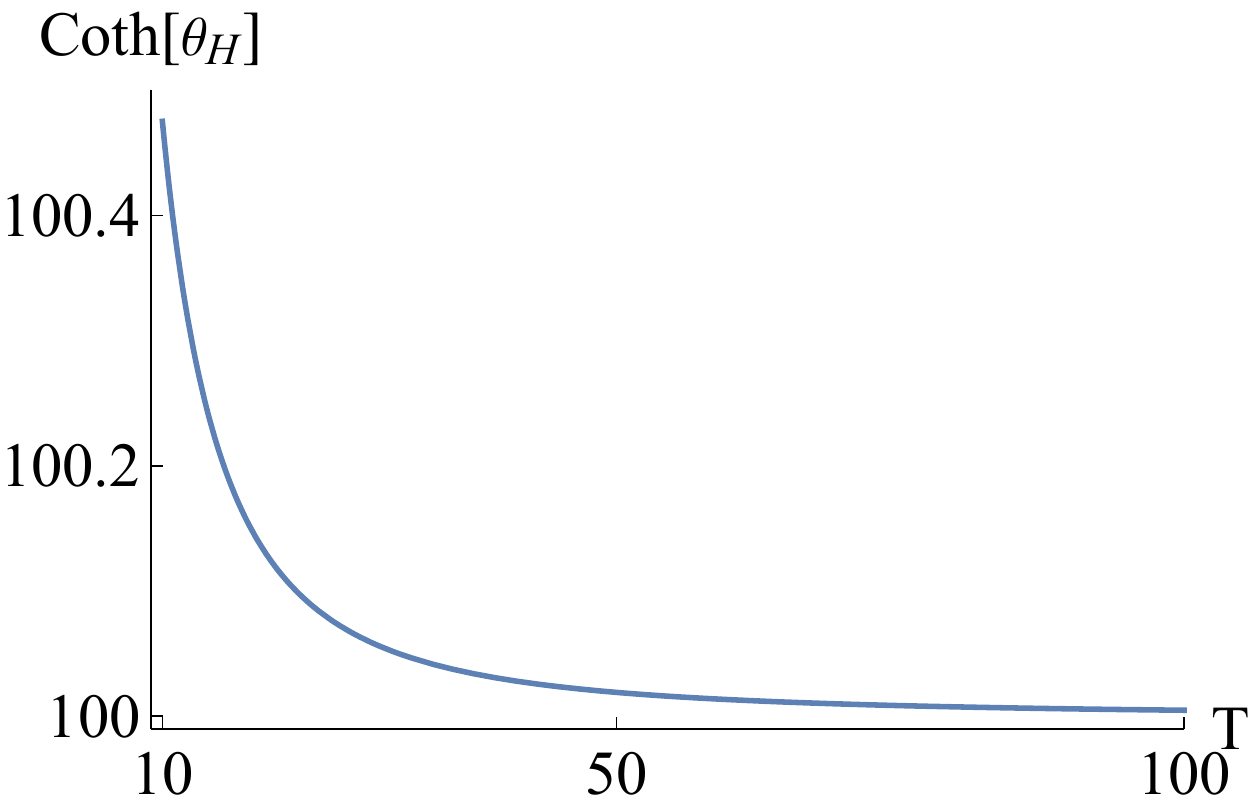}\ \hspace{0.8cm}
\includegraphics[scale=0.5]{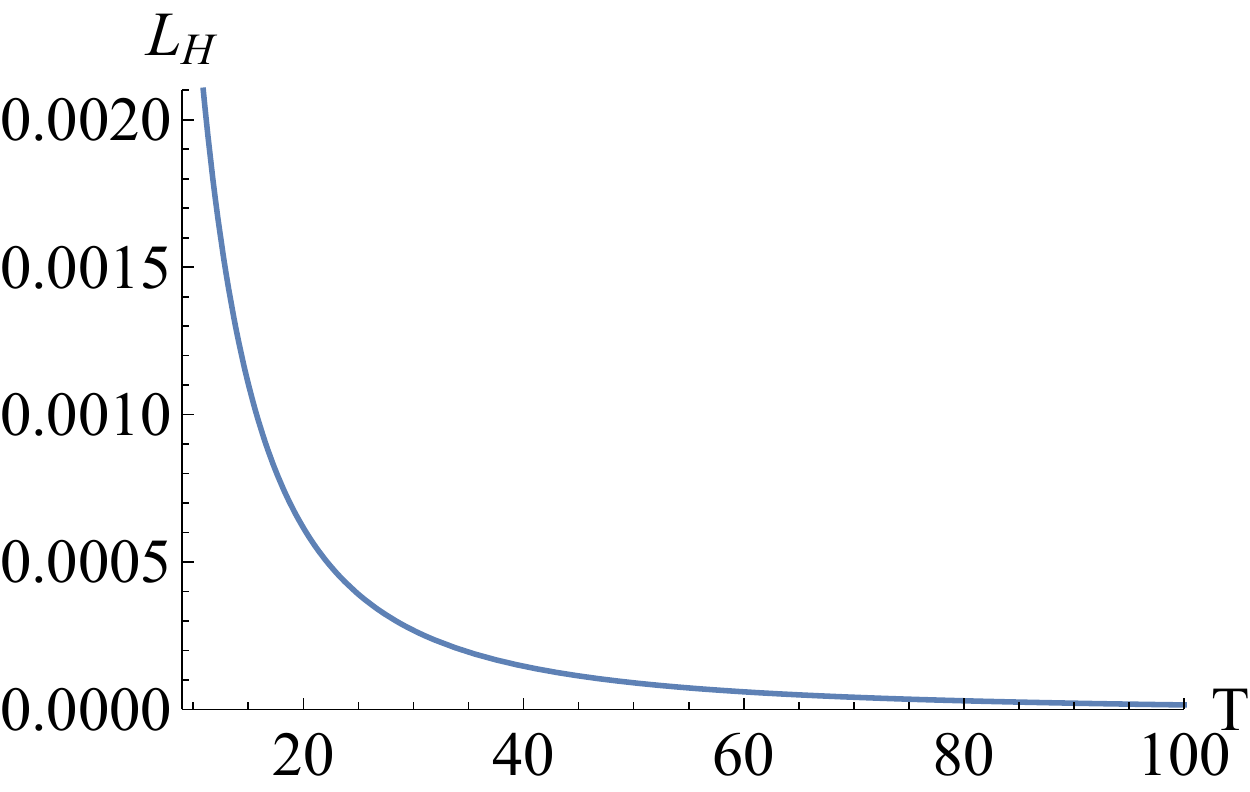}\ \hspace{0.8cm}\ \\
\includegraphics[scale=0.5]{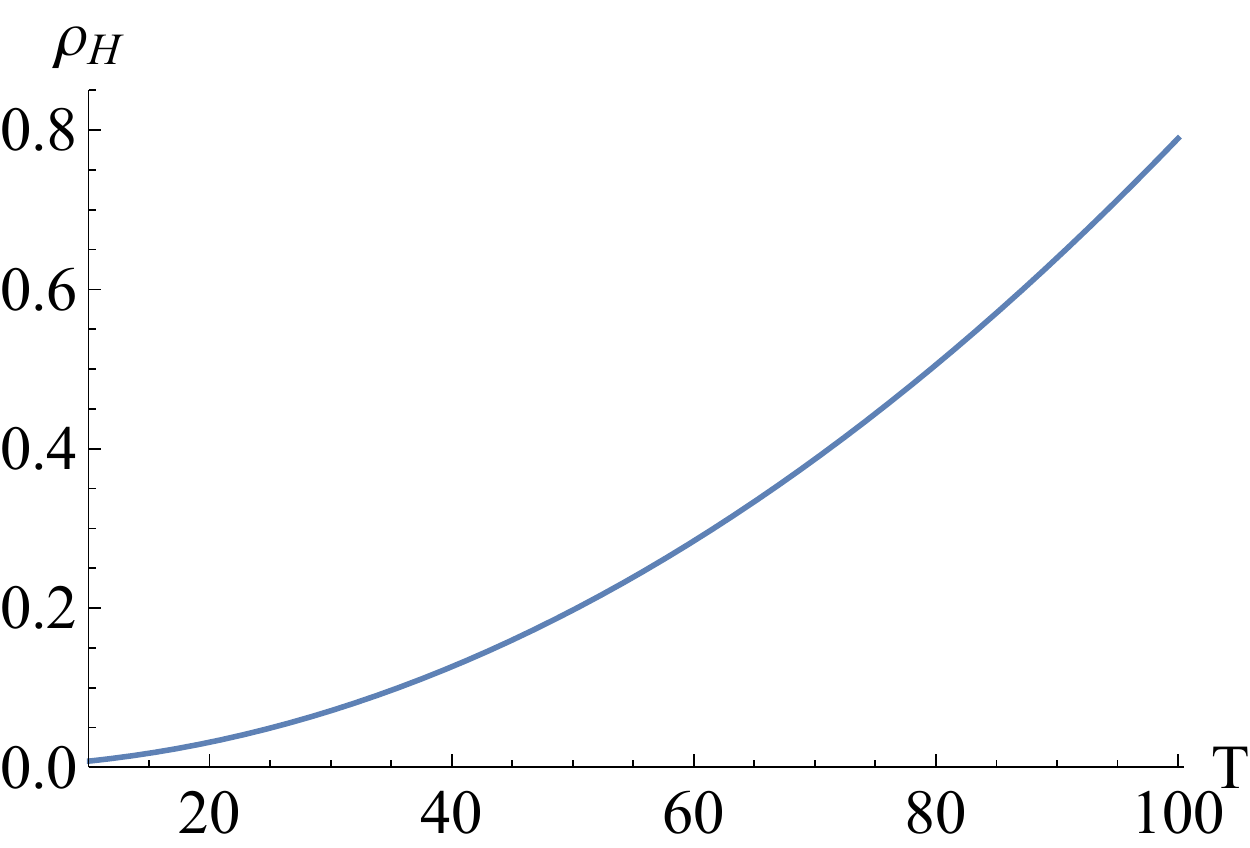}\ \hspace{0.8cm}
\includegraphics[scale=0.5]{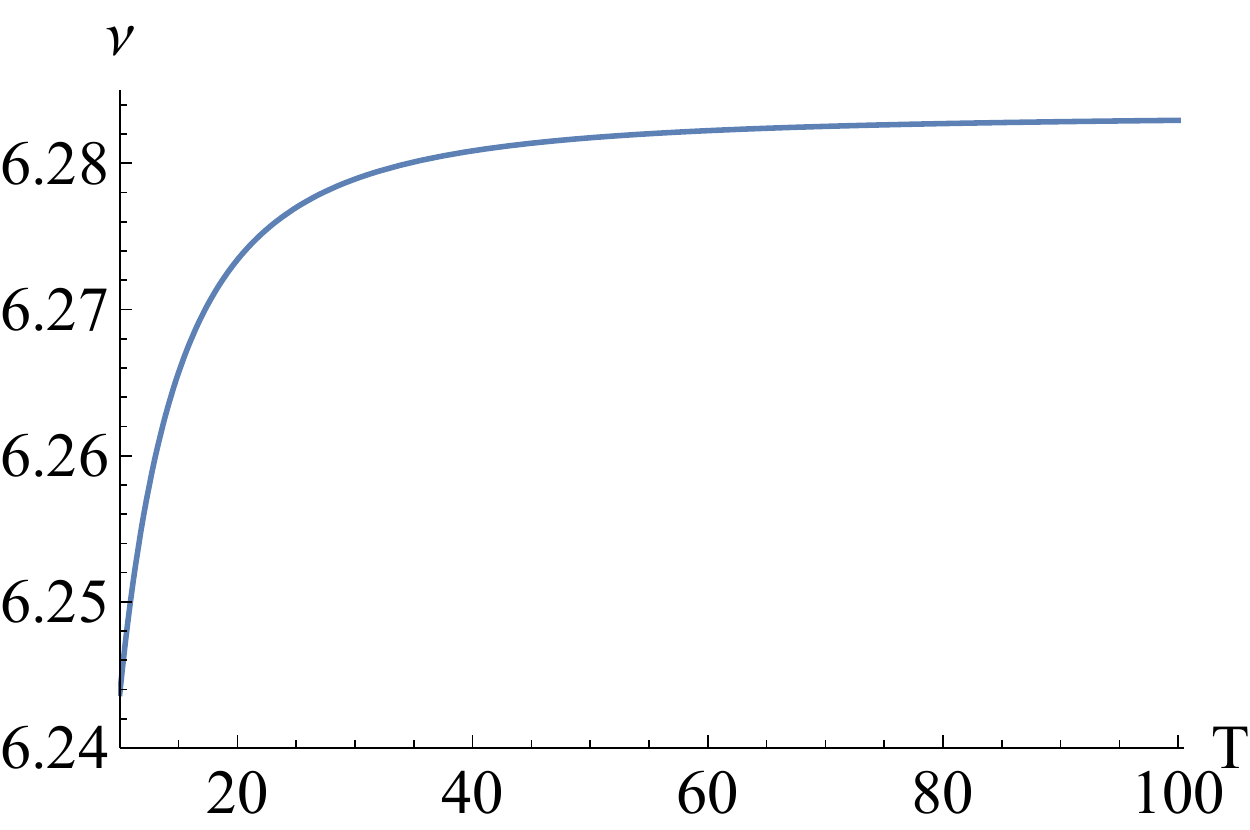}\ \hspace{0.8cm}\ \\
\caption{\label{tra-htem} The transports as the function of
temperature $T$ in the high temperature region. Here we have
set $k=1$, $q=10$ and $B=1$.}}
\end{figure}

The characteristics of the transport behavior in the high
temperature region are summarized as what follows.
\begin{itemize}
  \item Both DC resistivity $\rho_{dc}$ and thermopower $S$ decrease with $1/T^2$ at high temperature, which implies the thermal transport is dominant over the
electric and electrothermal transport.
  \item With the increase of temperature, the Hall angle $1/\tan\theta_H$ and Hall Lorentz ratio decrease (see the plots in FIG.\ref{tra-htem}),
while the magneticresistance $\rho_{B}$ increases (see the left
below plot in FIG.\ref{tra-htem}). The Nernst coefficient $\nu$
becomes a constant in the limit of high temperature (see the
right below plot in FIG.\ref{tra-htem}).
\end{itemize}

\section{Conclusions and discussions}
In this paper we have constructed a new class of charged black
brane solutions in EMDA theory, which is characterized by two free
parameters $a,\,b$, which could be viewed as the extension of
various charged solutions with $b=0$ in
literature \cite{Gubser:2009qt,Gao:2005xv,Gao:2004tv,Gouteraux:2014hca,Caldarelli:2016nni}.

For different $a,b$, the background exhibits distinct
behavior in zero temperature limit. In the neutral background $q=0$, the zero temperature ground state with
zero entropy density always exists for any $a,b$, while in the simple charged case $q\neq0,b=0$, it depends on $a$.
For $a^2<1/3$, the zero
temperature can be achieved at finite horizon position. But the
entropy density is finite as well at the zero temperature. It is
interesting to notice that a ground state with vanishing entropy
density is allowed for $a^2=1/3$, which has previously been
obtained in Gubser-Rocha model, and for $a^2>3$. While for $3>
a^2>1/3$, the zero temperature can not be achieved and the deep
horzion $z_+\gg1$ corresponds to a high temperature limit.
When the translation invariance is broken by adding axion
fields, by contrast,
the vanishing entropy density ground state can be achieved only
for $a^2=1/3$. For this special case we have demonstrated that the
dual system is characterized by a linear-T resistivity in a large
range of temperature, reminiscent of the key feature of the
strange metal. We have also obtained dyonic black brane by virtue
of the EM duality for $a=1$. The transport coefficients have been
calculated and their temperature dependence have been analyzed in
high temperature region. We expect the EM duality as a
valuable strategy may be applicable to more general gravity
theories such that more analytic solutions of dyonic black brane
could be constructed, which should be helpful for us to
investigate the magnetic transport behavior of the dual system by
holography.

\begin{acknowledgments}

We are very grateful to Chao Niu for many useful discussions and
comments on the manuscript. This work is supported by the Natural
Science Foundation of China under Grant Nos. 11575195, 11775036,
11747038, 11875053, 11847313.

\end{acknowledgments}

\begin{appendix}

\section{Electrically, magnetically charged dilatonic black branes and EM duality}\label{sec-EM-duality-bsol}
The well-known dyonic black brane solution is given in
Einstein-Maxwell theory, which enjoys S-duality. While, usually it
is hard to obtain new analytical solutions for dyonic black
brane. In this section, we shall study the electrically,
magnetically charged black brane solutions in EMD theory
\eqref{EMdac} with the help of EM duality.

\subsection{Pure magnetic solution}

We first construct the purely magnetic solution. For this
purpose, we introduce EM duality of the EMD theory \eqref{EMdac}.
We define the dual field strength $G_{ab}$ by the Hodge star
operation
\begin{equation}
G_{ab}:=\frac{Z(\phi)}{2}\epsilon_{abcd}F^{cd}\,,\label{ddrel}
\end{equation}
where $G=dH$ with $H_a$ being the dual gauge field and
$\epsilon_{abcd}$ is the completely antisymmetric Levi-Civita tensor.
And then, we can write down the dual one of EMD theory as
\begin{eqnarray}
\hat{S}_{EMD}=\int d^4x \sqrt{-g}\Big(R-\frac{1}{4Z(\phi)}G_{ab}G^{ab}-\frac{1}{2}(\partial \phi)^2+V(\phi)-\sum_{I=x,y}(\partial\psi_I)^2\Big)\,.\label{dEMdac}
\end{eqnarray}
From Eqs. \eqref{ddrel} and \eqref{dEMdac}, it is easy to find
that under the EM duality, the gauge coupling transforms as
$Z(\phi)\rightarrow 1/Z(\phi)$, which also implies a weak-strong
coupling duality. Especially, there is a correspondence between
the electric field $F_{tx}$ of the original theory and the
magnetic field $G_{yz}$ of its dual one. Therefore, by EM duality
we could quickly obtain a purely magnetic solution of the
dual EMD theory \eqref{dEMdac}. We demonstrate it as follows.

Given an electrically charged solution for the action in
\eqref{EMdac} which has a gauge coupling with an exponential
function $Z(\phi)=e^{a\phi}$, one can obtain a purely magnetic
solution for the dual action in \eqref{dEMdac} with gauge coupling
$1/Z(\phi)=e^{-a\phi}$ by virtue of the EM-duality, namely replace
$q\rightarrow B$ in \eqref{Gsol2}. Next, we change $a\rightarrow-a$ in the magnetic
solution. The final result is a solution for the action with
gauge coupling $Z(\phi)$ and potential
$\tilde{V}(\phi)=V(a\rightarrow-a,\phi)$, which is
\begin{eqnarray}
\hat{\hat{S}}_{EMD}=\int d^4x
\sqrt{-g}\Big(R-\frac{e^{a\phi}}{4}F^2-\frac{1}{2}(\partial
\phi)^2+\tilde{V}(\phi)-\sum_{I=x,y}(\partial\psi_I)^2\Big)\,.
\end{eqnarray}
One can obtain the following purely magnetic solution
\begin{subequations}
\label{sol-dual}
\begin{align}
&ds^2=\frac{1}{z^2}\Big(-\tilde{f}(z)dt^2+\frac{dz^2}{\tilde{f}(z)}+(1+\Lambda z)^{\frac{2a^2}{1+a^2}}(dx^2+dy^2)\Big)\,,
\label{sol-dual-metric}\\
&\tilde{f}(z)=(1+\Lambda z)^{\frac{2a^2}{1+a^2}}h(z)\,,\\
&h(z)=1+\frac{B^2(1+a^2)}{4\Lambda}z^3(1+\Lambda z)^{\frac{1-3a^2}{1+a^2}}-k^2z^2(1+\Lambda z)^{\frac{1-3a^2}{1+a^2}}\nonumber\\
&\qquad-b\Big(\frac{(1+\Lambda z)^{\frac{-3a^2+1}{a^2+1}}-1}{3a^2-1}+\frac{(1+\Lambda z)^{\frac{-a^2+3}{a^2+1}}-1}{a^2-3}-\frac{(1+\Lambda z)^{\frac{-2a^2+2}{a^2+1}}-1}{a^2-1}\Big)\,,\label{sol-dual-f}
\\
&\phi=\frac{2a}{1+a^2}\ln(\Lambda z+1)\,,\qquad \tilde{A}=Bxdy\,,\label{sol-dual-phi}
\end{align}
\end{subequations}
Note that this magnetic solution has also been found in \cite{Caldarelli:2016nni}.

\subsection{EM S-duality and dyonic black brane solution}

In this subsection, we construct the dyonic black brane solution from the EMDA theory \eqref{EMdac} for $a^2=1$,
in which the theory \eqref{EMdac} is S-duality, namely the charge and its dual magnetic solutions are both valid for the same action.

For this case, the potential becomes
\begin{subequations}
\begin{align}
&V=V_0(1+\Phi)+V_1\,,\\
&V_0=\frac{3(e^{\frac{1}{2}\phi}+ e^{-\frac{1}{2}\phi})^2-(e^{\frac{1}{2}\phi}-e^{-\frac{1}{2}\phi})^2}{2}\,,\\
&V_1=\frac{b}{2}(e^\phi-1)^3(e^{-\phi}+e^{-2\phi})\,,\\
&\Phi=-b\Big(\sinh \phi-\phi\Big)\,,
\end{align}
\end{subequations}
the charged black brane solution
\eqref{Gsol1} becomes
\begin{subequations}
\label{sol-sd}
\begin{align}
&
ds^2=\frac{1}{z^2}\Big(-f(z)dt^2+\frac{dz^2}{f(z)}+(1+\Lambda z)(dx^2+dy^2)\Big)\,,\label{sol-sd-ds}
\\
&f=(1+\Lambda z)h(z)\,,
\label{sol-sd-f}
\\
&h(z)=1-\frac{q^2z^3}{2\Lambda(1+\Lambda z)^2}-\frac{k^2z^2}{(1+\Lambda z)}-b\left(\frac{\Lambda  z (\Lambda  z+2)}{2 \Lambda  z+2}-\log (1+\Lambda z)\right)\,,\\
&A=A_t(z)dt=\Big(\mu-\frac{qz}{1+\Lambda z}\Big)dt\,,
~~~~~~\phi=\ln(\Lambda z+1)\,.
\label{sol-sd-At}
\end{align}
\end{subequations}

Since for $a^2=1$, the theory is S-duality, we also has the magnetic black brane solution
\begin{subequations}
\label{sol-magnetic-a-1}
\begin{align}
&\tilde{f}=(1+\Lambda z)h(z)\,,\\
&h(z)=1+\frac{B^2z^3}{2\Lambda(1+\Lambda z)}-\frac{k^2z^2}{(1+\Lambda z)}-b\left(\frac{\Lambda  z (\Lambda  z+2)}{2 \Lambda  z+2}-\log (1+\Lambda z)\right)\,,\\
&\tilde{A}=Bxdy\,,
~~~~~\phi=\ln(\Lambda z+1)\,.
\end{align}
\end{subequations}
Combining the charge black brane solution and the magnetic one,
one can easily construct the dyonic black brane solution
from the EMDA theory \eqref{EMdac}, which is
\begin{subequations}
\label{sol-EM-a-1}
\begin{align}
&f=(1+\Lambda z)\Big(1-\frac{q^2}{2\Lambda}z^3(1+\Lambda z)^{-2}+\frac{B^2}{2\Lambda}z^3(1+\Lambda z)^{-1}-b(\frac{\Lambda  z (\Lambda  z+2)}{2 \Lambda  z+2}-\log (1+\Lambda z))\Big)\,,\\
&A=\Big(\mu-\frac{qz}{1+\Lambda z}\Big)dt+Bxdy\,,~~~~~~\phi=\ln(\Lambda z+1)\,.
\end{align}
\end{subequations}
The line element is also \eqref{sol-sd-ds}.

\end{appendix}

\end{document}